\documentclass[prd, aps, superscriptaddress, preprintnumbers, twocolumn, floatfix, nofootinbib]{revtex4}

\usepackage{amsfonts}
\usepackage{amsmath}
\usepackage{amssymb}
\usepackage{bm}
\usepackage{dcolumn}
\usepackage{graphicx}   
\usepackage[latin1]{inputenc}
\usepackage{latexsym}
\usepackage{rotating}
\usepackage{graphicx}
\usepackage{color}
\usepackage{float}
\usepackage{epsfig}

\usepackage[unicode=true,pdfusetitle,
 bookmarks=true,bookmarksnumbered=false,bookmarksopen=false,
 breaklinks=false,pdfborder={0 0 1},backref=false,colorlinks=true]
 {hyperref}
\hypersetup{
 linkcolor=blue, citecolor=magenta, urlcolor=red, filecolor=blue}

\newcommand\be{\begin{equation}}
\newcommand\bea{\begin{eqnarray}}
\newcommand\ee{\end{equation}}
\newcommand\eea{\end{eqnarray}}

\begin{document}

\title{Dual Space-Time and Nonsingular String Cosmology}

\author{Robert Brandenberger}
\email{rhb@physics.mcgill.ca}
\affiliation{Physics Department, McGill University, Montreal, QC, H3A 2T8, Canada}

\author{Renato Costa}
\email{Renato.Santos@uct.ac.za}
\affiliation{Cosmology and Gravity Group, Dept. of Mathematics and Applied Mathematics,
University of Cape Town, Rondebosch 7700, South Africa}

\author{Guilherme Franzmann}
\email{guilherme.franzmann@mail.mcgill.ca}
\affiliation{Physics Department, McGill University, Montreal, QC, H3A 2T8, Canada}

\author{Amanda Weltman}
\email{awelti@gmail.com}
\affiliation{Cosmology and Gravity Group, Dept. of Mathematics and Applied Mathematics,
University of Cape Town, Rondebosch 7700, South Africa}

\date{\today}

\begin{abstract}

Making use of the T-duality symmetry of superstring theory, and of the double geometry from Double Field Theory, we 
argue that cosmological singularities of a homogeneous
and isotropic universe disappear. In fact, an apparent big bang singularity in Einstein gravity corresponds to a universe expanding to infinite size in the dual dimensions.

\end{abstract}

\maketitle

\section{Introduction} 

The singularities which arise at the beginning of time in both standard and inflationary
cosmology indicate that the theories which are being used in cosmology break down as
the singularity is approached. If space-time is described by Einstein gravity and matter
obeys energy conditions which are natural from the point of view of point particle theories,
then singularities in homogeneous and isotropic cosmology are unavoidable \cite{Hawking}.
These theorems in fact extend to inflationary cosmology \cite{Borde1, Borde2, Yoshida}.

But we know that Einstein gravity coupled to point particle matter cannot be the correct
description of nature. The quantum structure of matter is not consistent with a classical
description of space-time. The early universe needs to be described by a theory which
can unify space-time and matter at a quantum level. Superstring theory (see e.g.
\cite{Pol1, Pol2} for a detailed overview) is a promising candidate for a quantum
theory of all four forces of nature. At least at the string perturbative level, the building
blocks of string theory are fundamental strings. Strings have degrees of freedom and
new symmetries which point particle theories do not have, and these features may
lead to a radically different picture of the very early universe, as discussed many years
ago in \cite{BV} (see also \cite{other}).

As discussed in \cite{BV}, string thermodynamic considerations indicate that the
the cosmological evolution in the context of string theory should be nonsingular.
A key realization is that the temperature of a gas of closed string in thermal
equilibrium cannot exceed a limiting value, the {\it Hagedorn temperature} \cite{Hagedorn}.
In fact, as reviewed in the following section, the temperature of a gas of closed strings
in a box of radius $R$ decreases as $R$ becomes much smaller than the string
length. If the entropy of the string gas is large, then the range of values of $R$ for
which the temperature is close to the Hagedorn temperature $T_H$ is large. This
is called the {\it Hagedorn phase} of string cosmology. The exit from the Hagedorn
phase is smooth and is a consequence of the decay of string winding modes into
string loops\footnote{This mechanism suggests that exactly three spatial dimensions can
become large \cite{BV}, the others being confined to the string length by the
interaction of the string winding and momentum modes \cite{moduli}.}. The
transition leads directly to the radiation phase of Strandard Big Bang cosmology
(see \cite{SGCrevs} for reviews of the String Gas Cosmology scenario).

If strings in the Hagedorn phase are in thermal equilibrium, then the thermal fluctuations 
of the energy-momentum tensor can be computed using the methods of \cite{Deo}. 
In particular, it can be shown that in a compact space with stable winding modes
the specific heat capacity has holographic scaling as a function of the radius of the
volume being considered. As a consequence \cite{NBV, BNPV1}, thermal fluctuations of strings in
the Hagedorn phase lead to a scale-invariant spectrum of cosmological perturbations at
late times, with a slight red tilt like what is predicted \cite{Mukh} in inflationary cosmology.
If the string scale is comparable to the scale of particle physics
Grand Unification the predicted amplitude of the fluctuations matches the
observations well (see \cite{Ade:2015lrj} for recent observational results).
Hence, String Gas Cosmology provides an alternative to cosmological inflation as a theory for
the origin of structure in the Universe.
The predicted spectrum of gravitational waves \cite{BNPV2} is also scale-invariant, but
a slight blue tilt is predicted, in contrast to the prediction in standard inflationary
cosmology. This is a prediction by means
of which the scenario can be distinguished from standard inflation (meaning
inflation in Einstein gravity driven by a matter field obeying the usual energy
conditions). A simple modelling of the transition between the
Hagedorn phase and the radiation phase leads to a running of the spectrum
which is parametrically larger than what is obtained in simple inflationary
models \cite{Liang}. 

In this paper, we will study the cosmological background dynamics which follow
from string theory if the target space has stable winding modes. An example where
this is the case is a spatial torus. We will argue that from the point of view of
string theory the dynamics is non-singular.

\section{Dual Space from T-duality}

For simplicity let us assume that space is toroidal with $d = 9$ spatial dimensions, all of
radius $R$. Closed strings then have {\it momentum modes} whose energies are quantized
in units of $1/R$
\be \label{momentum}
E_n \, = \, \frac{n}{R} \, ,
\ee
where $n$ is an integer. They also have {\it winding modes} whose energies are quantized
in units of $R$, i.e.
\be \label{winding}
E_m \, = \, m R \, ,
\ee
where $m$ is an integer and we are working in units where the string length is one.
Strings also have a tower of {\ \it oscillatory modes} whose energies are independent of
$R$. The number of oscillatory modes increases exponentially with energy.

It follows from (\ref{momentum}) and (\ref{winding}) that the spectrum of string states
is invariant under the T-duality transformation
\be \label{tdual}
R \, \rightarrow \, \frac{1}{R}
\ee
if the momentum and winding numbers are interchanged. The transformation (\ref{tdual}) is
also a symmetry of the string interactions, and is assumed to be a symmetry of string theory
beyond perturbation theory (see e.g. \cite{Pol2})\footnote{See also \cite{Boehm} for an
extended discussion of T-duality when branes are added.}.

As is well known, the position eigenstates $|x \rangle$ are dual to momentum eigenstates $|p \rangle$.
In a compact space, the momenta are discrete, labelled by integers $n$, and hence
\be
|x \rangle\, = \, \sum_n e^{i n x} |n \rangle \, .
\ee
where $|n \rangle$ is the momentum eigenstate with momentum quantum number $n$.
As already discussed in \cite{BV}, in our string theory setting, windings are T-dual to momenta, and we can define a T-dual position operator
\be
|{\tilde x} \rangle \, = \, \sum_m e^{i m {\tilde x}} |m \rangle \, ,
\ee
where $|m \rangle$ are the eigenstates of winding, labelled by an integer $m$.

As again argued in \cite{BV}, experimentalists will measure physical length in
terms of the position operators which are the lightest. Thus, for $R > 1$ (in
string units), it is the regular position operators $|x \rangle$ which determine physical
length, whereas for $R < 1$ it is the dual variables $|{\tilde x} \rangle$. Hence, the
physical length $l_p(R)$ is given by
\begin{eqnarray} \label{plength}
l(R) \, &=& \, R  \,\,\,\, {\rm for} \,\,\,\, R \gg 1 \, ,\\
l(R) \, &=& \, \frac{1}{R} \,\,\,\, {\rm for} \,\,\,\, R \ll 1 \,  .  \nonumber
\end{eqnarray}

As was argued in \cite{BV}, in String Gas Cosmology the temperature singularity of the Big Bang
is automatically resolved. If we imagine the radius $R(t)$ decreasing from some initially
very large value (large compared to the string length), and matter is taken to be a gas of
superstrings, then the temperature $T$ will initially increase, since for large values of $R$ most
of the energy of the system is in the light modes, which are the momentum modes, and
the energy of these modes increases as $R$ decreases. Before $T$ reaches the maximal
temperature $T_H$, the increase in $T$ levels off since the energy can now go into producing
oscillatory modes. For $R < 1$ (in string units) the energy will flow into the winding modes which
are now the light modes. Hence,
\begin{equation} \label{Tdual}
T(R) \, = \, T\left(\frac{1}{R}\right) \, .
\end{equation}
A sketch of the temperature evolution as a function of $R$ is shown in Figure 1.
As a function of $\ln{R}$ the curve is symmetric as a reflection of the symmetry
(\ref{Tdual}). The region of $R$ when the temperature is close to $T_H$ and the
curve in Fig. 1 is approximately horizontal is called the ``Hagedorn phase''. Its
extent is determined by the total entropy of the system \cite{BV}.

\begin{figure}[h]
    \centering
    \includegraphics[scale = 0.4] {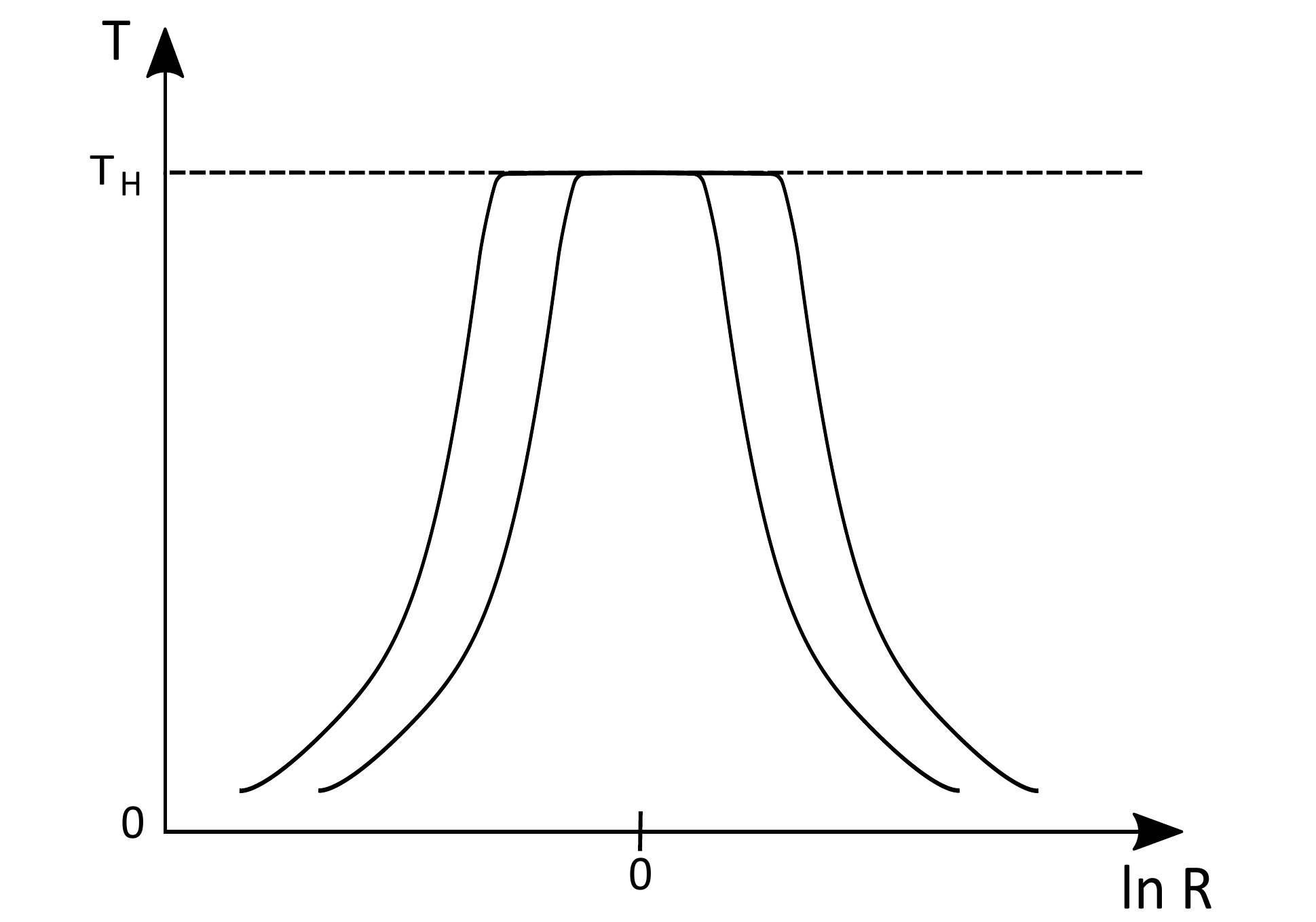}
    \caption{T versus $\log{R}$ for type II superstrings. Different curves are obtained for different entropy values, which is fixed. The larger the entropy the larger the plateau, given by the Hagedorn temperature. For $R=1$ we have the self-dual point.}
    \label{fig:my_label}
\end{figure}

\section{Cosmological Dynamics and Dual Space-Time}

In the following we will couple a gas of strings to a background appropriate
to string theory. Since the massless modes of string theory include, in addition
to the graviton, the dilaton and an antisymmetric tensor field, a
cosmological background will contain the metric, the dilaton and the
antisymmetric tensor field. For a homogeneous and isotropic cosmology
the metric can be written as
\be
ds^2 \, = \, - dt^2 + a(t)^2 d{\bf x}^2 \, ,
\ee
where $t$ is physical time, $a(t)$ is the cosmological scale factor and
${\bf x}$ are comoving spatial coordinates. We have assumed vanishing
spatial curvature for simplicity. We denote the dilaton by $\phi(t)$.

The T-duality symmetry of string theory leads to an important symmetry
of the massless background fields, the {\it scale factor duality} \cite{PBB}.
In the absence of an antisymmetric tensor field these take the form
\bea
a(t) \, &\rightarrow& \, \frac{1}{a} \\
{\bar \phi}(t) \, &\rightarrow& \, {\bar \phi}(t) \nonumber
\eea
where the T-duality invariant combination of the scale factor and the
dilaton is
\be
{\bar \phi} \, \equiv \, \phi - d  {\rm ln}a \, ,
\ee
where $d = D - 1$ is the number of spatial dimensions and $D$ the number of space-time dimensions.

The background equations of motion are those of dilaton gravity (we will neglect the 
antisymmetric tensor field). In the absence of matter, these equations
were studied in detailed in the context of Pre-Big-Bang cosmology \cite{PBB}.
In the presence of string matter, they have been analyzed in \cite{TV}. The
equations in the presence of a gas of matter described by energy
density $\rho$ and pressure $p$ are
\begin{align}
\left(\dot{\phi}-dH\right)^{2}-dH^{2} & =e^{\phi}\rho\label{eq:SUGRA_Hagedorn_1}\\
\dot{H}-H\left(\dot{\phi}-dH\right) & =\frac{1}{2}e^{\phi}p\label{eq:SUGRA_Hagedorn_2}\\
2\left(\ddot{\phi}-d\dot{H}\right)-\left(\dot{\phi}-dH\right)^{2}-dH^{2} & =0 \, ,
\label{eq:SUGRA_Hagedorn_3}
\end{align}
where $H\equiv \dot{a}/a$.
These are the equations in the string frame. In particular, we can combine these equations to write a continuity equation, 

\begin{equation}
    \dot{\rho} + (D-1)H(\rho + p) = 0.
\end{equation}

We consider matter to be a gas of strings. For $R \gg 1$ most of the energy is
in the momentum modes which act as radiation and hence have an
equation of state parameter $w \equiv p / \rho$ given by $w = 1/d$. For $R \ll 1$, however, most
of the energy density is in the winding modes whose equation of state parameter
is $w = - 1/d$. Finally, for $R = 1$ the equation of state is $w = 0$. An
interpolating form of the matter equation of state is
\begin{equation}
w\left(a\right) \, = \, \frac{2}{\pi d}\arctan\left(\beta {\rm ln} \left( \frac{a}{a_0} \right) \right),
\label{eq:SGmEOS}
\end{equation}
where $a_0$ is the value of the scale factor when $R = 1$, and $\beta$ is a constant
which depends on the total entropy of the gas. The larger the entropy is, the
wider the Hagedorn phase as a function of $a$, and hence the smaller the value
of $\beta$. For this equation of state, the continuity equation for string gas matter
can be integrated and yields
\bea
\ln\frac{\rho}{\rho_{0}} \, = \, &-& d \ln\frac{a}{a_{0}} -
\frac{2}{\pi}\left\{ \ln\left(\frac{a}{a_{0}}\right)\arctan\left[\beta\ln\left(\frac{a}{a_{0}}\right)\right] \right\}
\nonumber \\
&-& \frac{2}{\pi}\left\{ \frac{1}{2\beta}\ln\left[1+\beta^{2}\left(\ln\frac{a}{a_{0}}\right)^{2}\right]\right\} ,
\eea
where $\rho_{0}$ is the energy density at the string length. This result reproduces what is expected for large and small radii,
\bea
\rho\left(a\text{ large}\right) &\rightarrow& \rho_{0}\left(a/a_{0}\right)^{-(d + 1)} \\
\rho\left(a\text{ small}\right) &\rightarrow& \rho_{0}\left(a/a_{0}\right)^{-(d-1)}.
\eea
for pure momentum or pure winding modes, respectively.

At this point we have a system of background and matter in which both components
have the same symmetries. We now turn to an exploration of solutions. Following closely \cite{PBB}, we
make the ansatz
\bea
a(t) \, & \sim & \, \left( \frac{t}{t_0} \right)^{\alpha} \\
{\bar \phi}(t) \, & \sim & \, -\beta\ln \left(\frac{t}{t_0}\right) \, , \nonumber
\eea
where $\alpha$ and $\beta$ are constants, and $t_0$ is a reference time.
Inserting into the dilaton gravity equations gives the following constraints on
the constants
\bea \label{alg}
\left(D-1\right) w \alpha + \beta \, &=& \, 2 \\
\beta^{2} + \left(D-1\right)\alpha^{2} \, &=& \, 2 \beta \, . \nonumber
\eea

Deep in the Hagedorn phase when $w = 0$ we get
\be
( \alpha, \beta ) \, = \, ( 0, 2) \, .
\ee
This corresponds to a static scale factor in the string frame.
Converting to the Einstein frame in which the scale factor ${\tilde a}(t)$
is given by
\be
{\tilde a}(t) \, = \,  a(t) e^{- \phi/(d - 1)} 
\ee
we find
\be
{\tilde a}(t) \, \sim \, \left( \frac{t}{t_0} \right)^{2/(d - 1)} \, .
\ee

In the large $a$ phase when $w = 1/d$ we get
\be
( \alpha, \beta ) \, = \, \left( \frac{2}{D}, \frac{2}{D} (D - 1) \right) \, .
\ee
In this case, the dilaton is constant and hence the
string frame and Einstein frame scale factors are the
same. As expected, the scale factor evolves as in
a standard radiation dominated universe. There is a 
second solution of (\ref{alg}), but
that solution is consistent only for $p = 0$. 

When $w = - 1/d$ we have
\be
( \alpha, \beta ) \, = \, \left( - \frac{2}{D}, \frac{2}{D} (D - 1) \right) \, .
\ee
The string frame scale factor is expanding as we go
backwards in time. Translating to the Einstein frame
we get 
\be
{\tilde a}(t) \, \sim \,  \left( \frac{t}{t_0} \right)^{2/(d - 1)} \, .
\ee
In the Einstein frame, the scale factor vanishes at $t = 0$ while
in the string frame it blows up in this limit.

\begin{figure}[H]
    \centering
    \includegraphics[width=8
cm]{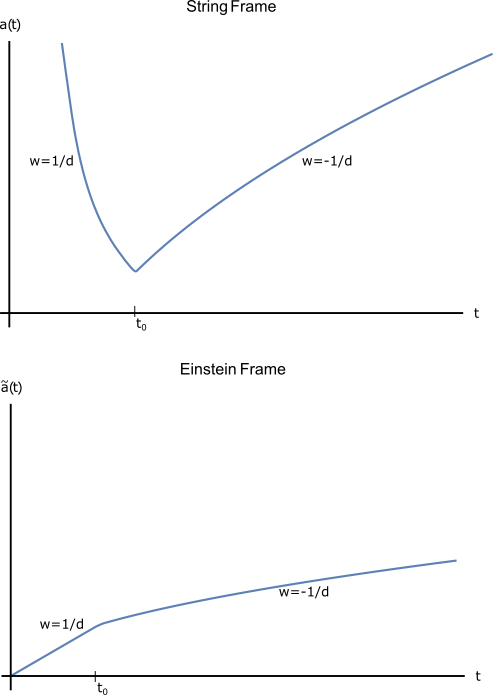}
    \caption{The schematic solution for the scale factor in the String and Einstein frames for $D=4$. Note that the transition between the winding and momenta equation of state has been smoothed out, as it is expected if (\ref{eq:SGmEOS}) is considered.}
    \label{fig3}
\end{figure}

Let us track the dynamics backwards in time, beginning with
a large torus ($R \gg 1$). The energy will hence be in
the momentum modes and the equation of state is that
of radiation. As we go back in time, the scale factor
decreases (it is the same in the two frames), 
the energy density increases, and eventually the temperature
approaches the Hagedorn value at which point oscillatory
and winding modes of the string gas get excited, leading
to a transition to an equation of state with $p =0$. We
enter a Hagedorn phase during which the string frame
scale factor is constant while the Einstein frame scale
factor is decreasing. This means that the radius of the
torus $R$ is decreasing, and it soon becomes energetically
preferable for the energy of the string gas to drift to
the winding modes, leading to an equation of state
$w = - 1/d$. In the winding phase the Einstein frame
scale factor is still decreasing, which is a self-consistency
check on the assumption that the energy of the string gas
is mostly in the winding modes\footnote{If we do not
allow momentum and winding modes to decay, then, as
studied in \cite{TV}, we obtain solutions where the string
frame scale factor oscillates about $a_0$.}.

We see that in the string frame, there is no curvature singularity.
As the coordinate time $t$ runs from $t = 0$ to $t = \infty$,
the scale factor is initially contracting, bounces in the
Hagedorn phase and expands afterwards in the radiation
phase, as showed schematically in Fig. 2.

Following \cite{us}, we argue that in the phase dominated by
winding modes we should measure time in terms of
the dual time variable
\be
t_d \, \equiv \,  - \frac{t_c^2}{t} \, \label{eq:PhysicalClockConstraint}
\ee
where $t_c$ corresponds to the coordinate time at the center
of the Hagedorn phase. In terms of $t_d$, the solution looks
like a contracting universe. 

From the point of view of the Einstein frame, the scale factor
vanishes at $t = 0$. But from the point of view of a detector
made up of winding modes, the measured scale factor is
proportional to $a(t)^{-1}$. Hence, the time interval
$0 < t < t_c$ corresponds to a contracting universe in terms
of the dual position basis.

Heuristically, there are two simple reasons for introducing a dual time coordinate. Let us consider for simplicity a fixed dilaton, so that we have a radiation solution. It is clear that there is an asymmetry between large and small scale factor, since the proper time for the scale factor to go to infinity diverges, while it is finite when the scale factor decreases to zero from some finite value. However, from the point of view of T-duality we should not be able to distinguish between a large and a small universe. This is the first hint towards a more general definition of the physical clock, $t_p$.

Another qualitative argument follows from special relativity considerations brought together with T-duality. For a large radius, rods are made out of momentum modes, and time measurements for a given physical length, $\Delta x$, are given by
\begin{equation}
\left|\Delta t\right|=\left|\Delta x\right|,\label{eq:zero_line_element-1}
\end{equation}
where the speed of light has been set to unit. If the universe is composed of closed strings, in principle we could have considered measuring physical length in terms of winding modes as well, and the natural rods built out of these modes are related to the physical length by 
\begin{equation}
\Delta\tilde{x} \rightarrow \frac{\alpha^{'2}}{\Delta x},
\end{equation}
where $\alpha^{'}$ is the string tension. Thus, we can rewrite (\ref{eq:zero_line_element-1})
as,
\begin{equation}
\left|\Delta\tilde{x}\right| \rightarrow \left|\frac{\alpha^{'2}}{\Delta t}\right|.
\end{equation}

Now, if we cannot distinguish large from small, we could have started the argument using winding modes instead, so that we would write the following relation\footnote{By T-duality one can argue that the dual speed of light is also equal to unit.},
\begin{equation}
\left|\Delta\tilde{x}\right|=\left|\Delta\tilde{t}\right|.
\end{equation}
Thus, it is also natural to propose a \emph{winding-clock}
that is dual to the momentum-clock by combining the above formulae,
\begin{equation}
\left|\Delta\tilde{t}\right| \rightarrow \left|\frac{\alpha^{'2}}{\Delta t}\right|.\label{eq:SRandTDuality}
\end{equation}

Evidently, physically speaking there is only a single clock. When
only winding or momentum modes are light, the existence of a unique
time coordinate is already clear. Around the self-dual point, when both modes are energetically favorable, that should also be the case. Therefore, we need
a prescription to reduce both time coordinates to a single
physical time. We call this prescription \emph{physical
clock constraint} and it is given by the identification (\ref{eq:PhysicalClockConstraint}). 

These ideas likely have a very natural interpretation in
terms of Double Field Theory \cite{DFT} (see also
\cite{DFTearly} for some early work). Double Field Theory
is a generalization of supergravity which lives in $2d$ spatial
dimensions, with the first $d$ dimensions corresponding to the
usual $x$ variables, and the second $d$ dimensions to the
dual spatial variables ${\tilde x}$. In Double Field Theory there
is a generalized metric which for homogeneous and isotropic
cosmology and in the absence of an antisymmetric tensor field
is given by
\begin{equation} \label{metricCOSMO}
ds^2 \, = \, - dt^2 + a^2(t) \delta_{ij} dx^i dx^j + a^{-2}(t)\delta^{ij} d\tilde{x}_i d\tilde{x}_j\, .
\end{equation}
The determinant of the generalized metric is one. As space
shrinks in the $x$ directions, it opens up in the ${\tilde x}$
directions. This is sketched in Fig. 3. 
In work in progress \cite{us2} we are exploring
this connection in more detail, in particular using the $O(D,D)$-formalism for formalizing the introduction of a dual time, and discussing how the physical clock constraint can be seen analogously to the imposition of the section condition in DFT for the dual coordinates \cite{Aldazabal:2013sca}. 

\begin{figure}
    \centering
    \includegraphics[width=8cm]{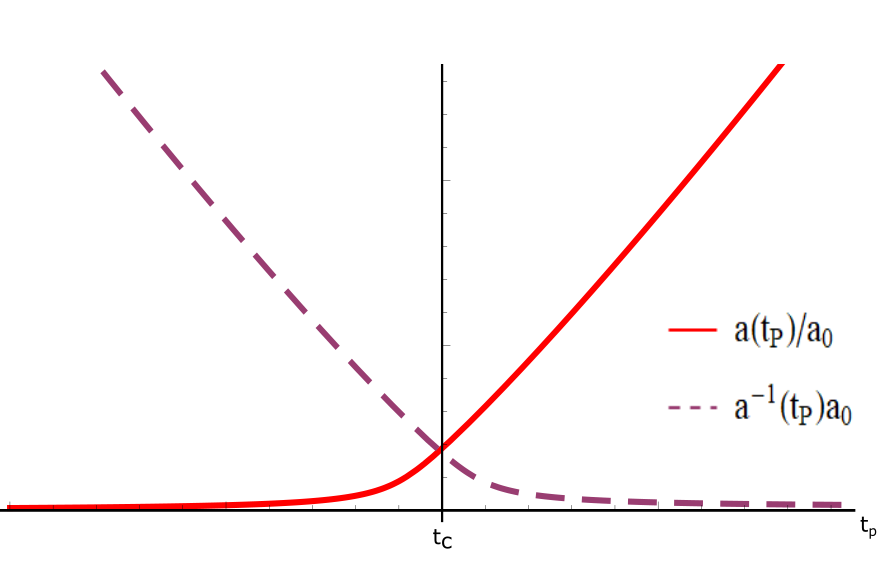}
    \caption{The scale factor goes to zero only at $t_p \rightarrow -\infty$. Similarly its inverse goes to zero when $t_p \rightarrow \infty$.}
    \label{fig2}
\end{figure}

\section{Discussion}

We have studied the equations of motion of a cosmological background
containing the scale factor $a(t)$ and the dilaton in the presence of string
gas matter sources. Both the background action and the matter action
are consistent with the T-duality symmetry of string theory. While we do
not expect our description to be adequate in the high density phase when
truly stringy effects must be considered, our analysis is an improvement
over the usual effective field theory of string cosmology where the 
underlying background geometry
is not covariant with the T-duality symmetry.

We find that the solutions are nonsingular, at least when interpreted in
the context of double space-time. We conjecture that an improved
description could be obtained using the tools of Double Field Theory\footnote{For a recent paper exploring the required formalism see \cite{Park}.}.

\section*{Acknowledgement}

This research is supported by the IRC - South Africa - Canada Research Chairs
Mobility Initiative Grant No. 109684. The research at McGill is also supported in
part by funds from NSERC and from the Canada Research Chair program.
Two of us (RB and GF) wish to thank the Banff International Research
Station for hosting a very stimulating workshop ``String and M-theory Geometries: Double Field Theory,
Exceptional Field Theory and their Applications" during which some of the
ideas presented here were developed. GF acknowledges financial support from 
CNPq (Science Without Borders) and PBEEE/Quebec Merit Scholarship. 
GF also wishes to thank the University of Cape Town where most of this
work was developed. R.C. acknowledges financial support by the 
SARChI NRF grantholder. A. W. gratefully
acknowledges financial support from the Department
of Science and Technology and South African
Research Chairs Initiative of the NRF.

\end{document}